\documentclass[12pt,preprint]{aastex}

\newcommand{\xtej}{XTE\,J1550$-$564}
\newcommand{\gx}{GX\,339$-$4}

\shorttitle{Spectral and Timing Evolution of \xtej}
\shortauthors{Reilly et al.}

\begin{document}

\title{USA Observation of Spectral and Timing Evolution \\
During the 2000 Outburst of \xtej}

\author{
K. T. Reilly\altaffilmark{1},
E. D. Bloom,
W. Focke,
B. Giebels,
G. Godfrey,
P. M. Saz Parkinson,\\
G. Shabad
}
\affil{Stanford Linear Accelerator Center, Stanford University, 
Stanford, CA 94309}
\author{
P. S. Ray\altaffilmark{2},
R. M. Bandyopadhyay\altaffilmark{3},
K. S. Wood,
M. T. Wolff,\\
G. G. Fritz,
P. Hertz\altaffilmark{4},
M. P. Kowalski,
M. N. Lovellette, 
D. J. Yentis
}
\affil{E. O. Hulburt Center for Space Research, Naval Research
Laboratory, Washington, DC 20375}
\and
\author{Jeffrey D. Scargle}
\affil{Space Science Division, NASA/Ames Research Center, 
Moffett Field, CA 94305-1000}
\altaffiltext{1}{Kaice.Reilly@slac.stanford.edu}
\altaffiltext{2}{Paul.Ray@nrl.navy.mil}
\altaffiltext{3}{NRL/NRC Research Associate}
\altaffiltext{4}{Current address: NASA Headquarters, 300 E Street, SW, Washington, DC 20546-0001}

\begin{abstract}
We report on timing and spectral observations of the 2000 outburst of
\xtej\  made by the Unconventional Stellar Aspect (USA) Experiment on
board the {\em Advanced Research and Global Observation Satellite}
(ARGOS).  We observe a low-frequency quasi-periodic oscillation
(LFQPO) with a centroid frequency that tends to increase with
increasing flux and a fractional rms amplitude which is correlated
with the hardness ratio.  The evolution of the hardness ratio (4--16
keV/1--4 keV) with time and source flux is examined.  The
hardness-intensity diagram (HID) shows a cyclical movement in the
clockwise direction and possibly indicates the presence of two independent
accretion flows.  We observe a relationship between the USA
4--16~keV count rate and radio observations and discuss this in the
context of previously observed correlations between X-ray, radio,
optical and IR data.  We examine our results in the context of models
invoking two accretion flows: a thin disk and a hot sub-Keplerian flow.
\end{abstract}

\section{Introduction}

\xtej\ was first observed in 1998 September by the All Sky Monitor 
(ASM) on board the {\em Rossi X-ray Timing Explorer} (RXTE) when it
began an outburst lasting approximately 8 months \citep{smi98}.
\xtej\ began a second outburst on 2000 April 2 \citep{ms00}, lasting
approximately 2 months.  The source was detected for a third time in
2001 January but did not go into a full outburst
\citep{tsw+01,jbt01}.

Recent optical observations of \xtej\ have placed a lower limit of
$7.4 \pm 0.7 M_\sun$ on the mass of the compact object
\citep{ovm+01}. This mass places the compact object well above the
maximum mass for a stable neutron star and so provides compelling
evidence that \xtej\ contains a black hole.  During its 1998--1999 outburst,
apparent superluminal radio jets ($v > 2c$) were observed by
\citet{hch+01}; however, the jet's angle to the line of sight has not yet
been determined.  Radio and optical observations of the 2000 outburst
also show evidence of jet formation \citep{ckj+01,jbo+01}.

Complex timing and spectral behavior has been observed by RXTE in
\xtej\ during its two full outbursts. This behavior includes
detections of three classes of low-frequency quasi-periodic oscillations
(LFQPO) ($< 20$~Hz) and several
detections of high frequency QPOs (HFQPO) ($>100$~Hz)
\citep{rsm+01,mwh+01,ktr+01}. Further, color-color diagrams and
hardness intensity diagrams of the 1998--1999 outburst showed separate
spectral branches for each of the black hole states as well as
correlations with quasi-periodic oscillations and other timing behavior
\citep{hwb+01}.

During the 1998--1999 outburst, \xtej\ exhibited all four identified
black hole spectral states \citep{smr+00,hwv+00}.  In the 2000
outburst the source never achieved the high state (HS), going from an
initial low/hard state (LS) to an intermediate state (IS) 
or a very high state (VHS)
and returning to a final LS \citep{mwh+01}.  Although previous authors
have made a distinction between the IS and VHS, recent work on \xtej\
has suggested that the IS and VHS are actually the same state at
different X-ray flux levels \citep{hwv+00}.  Therefore, for the
remainder of this paper we will refer to this state as the IS. The
transition from the initial LS to the IS was made on 2000 April 26
(MJD 51660). The transition from the IS back to the LS occurred
sometime between May 13 (MJD 51677) and May 19 (MJD 51623)
\citep{ckj+01}.

In this {\em Letter}, we report on X-ray observations of the 2000
outburst of
\xtej\ made by the Unconventional Stellar Aspect (USA) Experiment on
the US Air Force {\em Advanced Research and Global Observation
Satellite} (ARGOS).  For a detailed description of the USA experiment
see \citet{rwf+99} and \citet{sha+00}. We present lightcurves and hardness
ratios and track a low frequency QPO which appears in the initial and
late stages of the outburst.

\section{Observations and Data Analysis}

\subsection{Lightcurves and Hardness Ratios}

The USA Experiment observed \xtej\ at the rate of 
2--8 times per day between 2000
April 14 (MJD 51648) and June 18 (MJD 51713).  For the present
investigation, 193 observations were used, from which we selected
$\sim 49$~ks of data obtained in low-background regions.  The data are
time tagged, having 32 $\mu$s time resolution, and cover an energy
range of approximately 1--17~keV in 16 pulse height analyzer (PHA) 
channels. In this
work we do not make use of the lowest (channel 0) and highest (channel
15) PHA channel. We refer to PHA channels 1--14 ($\sim 1-16$~keV) as
the total range.

\begin{figure}
\begin{center}
\resizebox{\textwidth}{!}{
\includegraphics[angle=270]{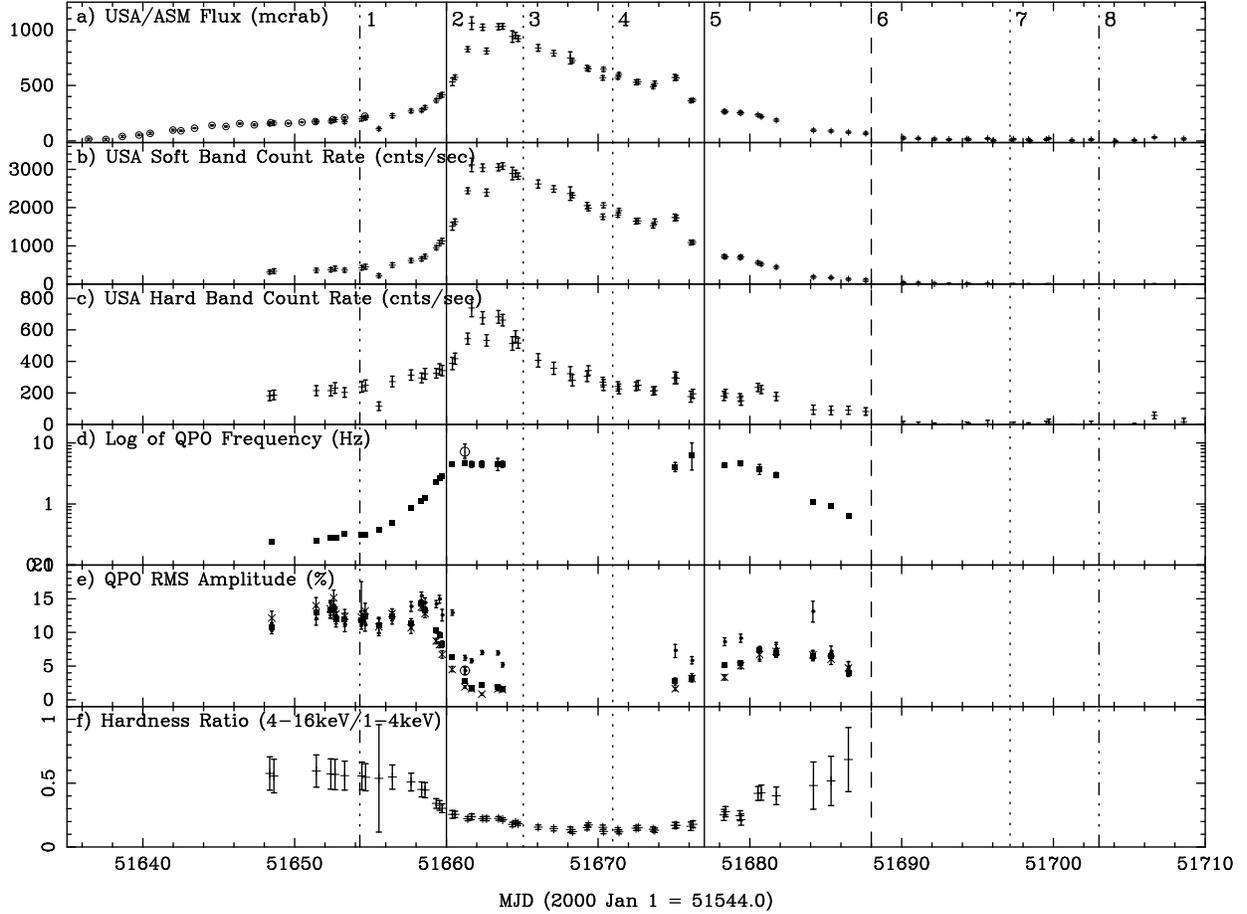}
}
\end{center}
\caption{Time Evolution of \xtej. 
(a) USA/ASM Crab normalized fluxes for the outburst. The circles are
ASM daily averaged data taken before USA started observations. ASM
error bars are shown but are smaller than the circles.  Crosses with
error bars are USA data. (b) USA Soft Band count rate (SB),
$\sim1-4$~keV.  (c) USA Hard Band count rate (HB), $\sim4-16$~keV.
(d) LFQPO centroid frequency.  Filled squares show results
for the energy range 1--16~keV.  The large unfilled circle shows the
frequency of the anomalous HB QPO.  (e) LFQPO
percent rms amplitude.  Filled squares show results for the energy
range 1--16~keV. Filled circles are the HB and the ``X''s are the SB.
The large unfilled circle marks the anomalous QPO.
(f) Hardness ratio (HB/SB). Line 2 shows the transition between the
initial LS and the IS. Line 5 shows the approximate beginning of the
transition period between the IS and final LS. Lines 3, 4 and 7 are
the times of ATCA observations. Lines 1 and 8 show optical/IR maxima
and line 6 shows an optical/IR minima.
\label{lightcurves}}
\end{figure}

To create the light curves shown in Figure~\ref{lightcurves}, a first
order background subtraction was made by averaging blank sky
observations and then subtracting these values from the count rate.  
To determine
the total error, the standard deviation of the average in the
background was added in quadrature to the error on the count rate. USA
data were then corrected for obscuration by the instrument support
structure and the collimator response.  The overall average light
curve (normalized to the USA Crab counting rate) for the total range
is shown in panel (a) of Figure~\ref{lightcurves}.  The circles are
RXTE/ASM daily averaged data used to give the complete outburst
profile (USA observations did not cover the first few days of the
outburst).  The USA data points shown are an average of several USA
observations.  The number of observations averaged was dependent on
the observation spacing and signal to noise ratio.

The spectral characteristics of the outburst were studied by dividing
the USA data into two energy bands, USA PHA channels 1--3 and channels
4--14.  These two bands correspond to 1--4~keV and 4--16~keV,
respectively.  For the remainder of this {\em Letter} these bands will
be referred to as the soft band count rate (SB) and the hard band
count rate (HB).  The
motivation for choosing the specific energy ranges of the SB and HB
came from properties of the hardness-intensity diagram (HID).  During
the outburst, \xtej\ traces a cyclic pattern in the HID.  Plots of
count rates versus the total range show this cyclical structure for
individual channels in the range 4--14, but not for channels 1, 2 or
3.

\begin{figure}
\resizebox{3.5in}{!}{
\includegraphics{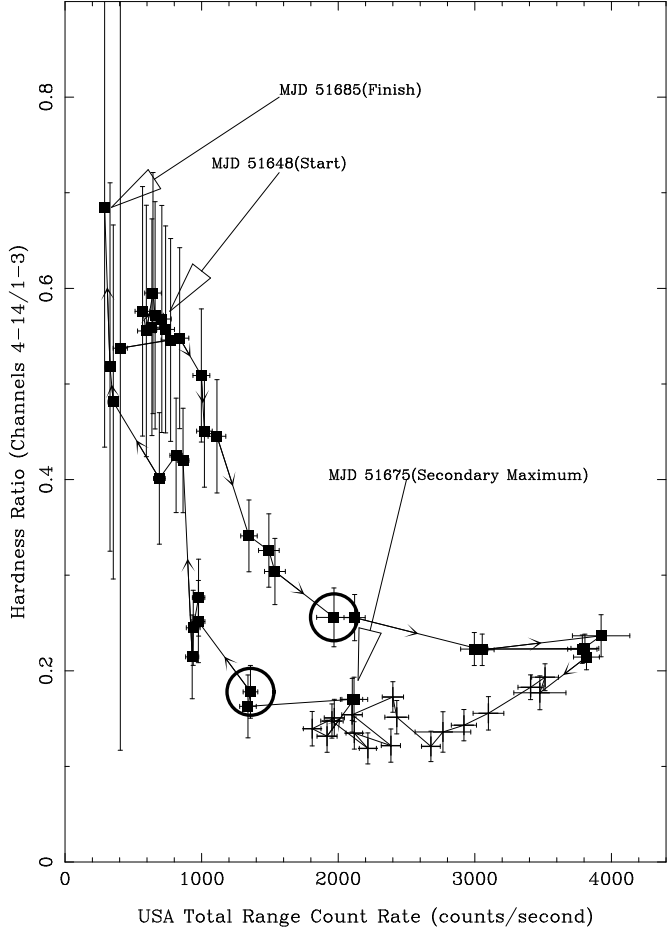}
}
\begin{center}
\caption{Hardness Ratio (HB/SB) vs. Total Range Count Rate.  Squares are
plotted on points in which the LFQPO was detected.  Points at which no
QPO detection was made only show error bars. The error bars shown are
the standard deviation of the flux and the hardness ratio.  Arrows
plotted between points show the direction of time. The larger outlined
arrows mark the start and stop times of USA observations and the time
of the secondary maximum.  Transitions between states have been marked
with the bold circles.
\label{HRintensity}}
\end{center}
\end{figure}

Panels (b) and (c) of Figure~\ref{lightcurves} show the SB and HB as a
function of time during the outburst. The bottom panel of
Figure~\ref{lightcurves} shows the evolution of hardness ratio using
these energy bands.  The HID in Figure~\ref{HRintensity} shows how the
hardness ratio evolves with total count rate during the outburst.  The
hardness ratio is plotted only for USA data prior to MJD 51687, after
which the signal to noise decreases to the point that the hardness
ratio is not constrained.  In Figure~\ref{HRintensity} and panel (f)
of Figure~\ref{lightcurves}, one point is shown for each point in
panels (b) and (c) of Figure~\ref{lightcurves}.

The design of the USA detector incorporated automatic gain
stabilization hardware and frequent iron source energy calibrations
were done while in orbit. We note that the USA channel to 
energy conversion varies slightly over the USA orbit;
however, checks
performed showed that this variation made no significant impact on the
relevant features seen in the SB, HB, and hardness ratio.
A further check of our spectral
results was made by comparing our data to public RXTE/ASM data.  Daily
averaged ASM data were used to find the hardness ratio as a function
of time and to make a HID. The ASM hardness ratio was calculated by
dividing the sum of ASM B Band (3--5~keV) and C Band (5--12~keV) by A
Band (1.5--3~keV).  To try to emulate the ASM energy bands the USA PHA
channels 3--11 (3--11.5~keV) were summed and divided by channel 2
(2--3~keV).  The ASM data confirmed the hardness ratio observed with
USA and the cyclic behavior in the HID.

\subsection{Power Spectra: Low Frequency QPOs}

A low frequency quasi-periodic oscillation (LFQPO) was ubiquitous
during the rise of the outburst and during the decay of the outburst
after the secondary maximum.  In order to track the LFQPO evolution
through the outburst, observations were grouped by day and frequently
in sub-day groups (signal to noise ratio permitting).

Power spectra (see \citealt{nvw+99} and references therein) were
calculated from these groups and averaged.  The resultant power
spectrum for each group containing a LFQPO was fit with a power law or
broken power law and a Lorentzian for any observable QPO features.
Fits were made in three energy bands: the total range, SB, and HB.  In
case of confusion by sub-harmonics, the strongest QPO feature was
chosen as the primary LFQPO (see \citealt{rsm+01}).  In most cases no
sub-harmonics were detectable.  Panel (d) of Figure~\ref{lightcurves}
shows how the centroid frequency of the LFQPO evolves during the
outburst.  The evolution of the rms amplitude for all three energy
bands is shown in panel (e).  Panels (d) and (e) show error bars
calculated by allowing the $\chi^{2}$ of the fit to vary by one.  All
error bars are given at the 68\% confidence level.

\section{Results}

\subsection{QPO Evolution and Correlation to State Changes}

We observe LFQPOs between MJD 51648 and 51663 and between MJD 51675
and 51686 which vary in frequency between 0.24--7.19~Hz and
6.34--0.64~Hz, respectively. During the times of these detections the
source is either in the LS or near the transition from one state to
another.  The LFQPO rms amplitude decreases rapidly at the state
transition from the LS to the IS and then increases during the
transition back to the LS, indicating that the mechanism for creating
the LFQPO is suppressed in the IS.  The LFQPO centroid frequency
generally increases with increasing flux; the fractional rms amplitude
is correlated with hardness ratio (see Figure~\ref{lightcurves}).

During the IS, significant detections of HFQPOs (249--278~Hz) were
made by RXTE between between MJD 51663 and 51675 \citep{mwh+01}. A
65~Hz QPO has been discovered by \citet{ktr+01} at MJD 51684.8.  These
HFQPO detections occurred during the periods where USA observed the
LFQPO to be weakening or not detectable at all.  It is interesting to
note that the HFQPOs were observed to decrease in strength as a
function of time in the IS \citep{mwh+01}. These observations point to
an unfavorable interaction between the mechanisms for LFQPO and HFQPO
production. We observe a decline in rms amplitude of the LFQPO near
the LS/IS transition, which marks the approximate onset of the
HFQPOs. Towards the end of the IS, near the IS/LS transition, the
HFQPO weakens as the LFQPO once again becomes detectable.  This trend
continues in the last days of the outburst, when the LFQPO rms
amplitude weakens and the 65~Hz QPO is detected.

These QPO features are qualitatively consistent with observations of
\xtej\ ~during the 1998--1999 outburst \citep{rsm+01}, 
during which an ``antagonism'' between LFQPOs and HFQPOs was also
observed.  During that outburst, type ``C'' QPOs were observed when
strong correlations were seen between the frequency and disk flux
while the amplitude was observed to correlate with disk temperature
\citep{rsm+01}.  These previous observations closely resemble what we
see for the LFQPO observed during the 2000 outburst; thus we
tentatively classify the LFQPO discussed here as a type C.

At MJD 51661.21 an anomalous QPO was detectable exclusively in the HB,
in contrast to the primary LFQPO which appears in all energy ranges.
In Figure~\ref{lightcurves} panel (d), this anomalous QPO is the
highest frequency point and is marked with a large unfilled circle
near MJD 51660. This QPO was detected between the primary LFQPO and
its harmonic and has a frequency of $7.19^{+0.12}_{-0.11}$~Hz, whereas
the primary LFQPO is seen at $4.71\pm0.05$~Hz and its harmonic is seen
at $9.75^{+0.4}_{-0.35}$~Hz.

\subsection{Spectral Evolution\label{spectev}}

From Figure~\ref{lightcurves} it is clear that the 2000 outburst of
\xtej\ does not follow the canonical fast rise, exponential decay
(FRED) outburst as would have been expected prior to the RXTE era.
Now, with many more examples of well-observed soft X-ray transient
(SXT) outbursts, it has become clear that few outbursts look like pure
FREDs, and that the outburst profile can be very different at high
energy than at low energy.  Comparing the SB and HB lightcurves in
Figure 1 (b) and (c), the two bands rise approximately in unison, but
show very different behavior after the peak.  With the exclusion of
the secondary maximum the HB light curve shows a nearly symmetric
outburst profile, while the SB rapidly rises, then decays
approximately linearly.  As is often observed in SXTs, the decay
returns to its original path after the secondary maximum.

In the HID (Figure~\ref{HRintensity}), we observe that the difference
between the SB and HB lightcurves manifests itself as a cyclic
structure that moves with time in a clockwise direction.  The HID
shows a rapid drop in the hardness ratio as the source enters the IS
and a rapid increase as the final LS is realized.  This type of
spectral structure, which has been seen in several other sources
\citep{shs+01}, is a consequence of the spectrum being harder during
the rise than during the decline.

\subsection{Multiwavelength Correlations}
\label{sec-multi}

Multiwavelength observations of \xtej\ during the 2000 outburst have
shown correlations between radio, optical, IR and X-rays.  These
observations have been compared with similar correlations seen in
other black-hole candidates (BHCs), such as \gx, and have led to the
interpretation that the LS of BHCs is characterized by jets and the IS
state is characterized by quenching of these jets \citep{ckj+01}. In
Figure~\ref{lightcurves}, lines 1 and 8 mark the times of optical/IR
maxima and lines 3 and 7 indicate significant radio detections made by
ATCA while line 4 indicates no significant radio detection. Line 6 is
the time of an optical/IR minimum. (Optical/IR observations were made
by YALO; for complete details on the radio and optical/IR observations
see \citealt{ckj+01} and \citealt{jbo+01}).  Radio observations by MOST
(not shown) made during the initial LS show a detection of 8--15~mJy
at 843~MHz \citep{ckj+01}. This indicates that the radio emission
prior to line 3 is more than a factor of 50 stronger than at the time
of the ATCA observation made at line 4 \citep{ckj+01}. Together, these
radio and optical/IR observations have been 
interpreted\citep{ckj+01,jbo+01} as evidence for the
presence of jets during the initial and final LS.  USA observations
indicate that the decline in the HB, during the IS, is associated with
a rapid drop in radio emission. The weak radio detection (line 3)
occurs while the SB is still very near its maximum and the HB has dropped
from its maximum by a factor of $\sim2$ and continues on a rapid
decline. At the point of no significant radio detection (line 4), the
HB has greatly reduced its rate of decrease. The return of the radio
signal in the final LS suggests that the mechanism of jet production
is related to but not dependent on HB photon production mechanism. In
future outbursts, it would be helpful to obtain a much more thorough
radio coverage in order to pin down the nature of the relationship
between the HB and jet quenching.

\section{Discussion: Two Flow Models}
\label{2flow}

The different behavior of the HB and SB lightcurves, together with the
lack of a strict correlation between the QPO frequency and source flux,
suggest a scenario involving a two component accretion flow.  Models
invoking two {\em independent} accretion flows have been described by
\citet{ct95} and \citet{shs+01}, while \citet{van01} presents a model 
where the two parameters are the instantaneous and time averaged
values of a {\em single} quantity.

Whether the two accretion flows are dependent or independent, there
will be a transition radius at which the Keplerian disk is disrupted
and forms a hot sub-Keplerian flow.  This inner flow can be an
advection-dominated accretion flow \citep{emn+97} or a postshock flow
interior to the radius where two independent flows interact
\citep{ct95}.  One may then consider models where the LFQPO is related
to the Keplerian orbital period at the transition radius
($r_\mathrm{tr}$, measured in units of the Schwartzchild radius)
between the thin disk and an inner flow.  The relationship between the
Keplerian orbital period and radius for a $7.5 M_\sun$ black hole is
$R = 195 R_\mathrm{Sch} P^{2/3}$, where $R_\mathrm{Sch}$ is the
Schwartzchild radius and $P$ is the orbital period in seconds.  Thus,
the LFQPO evolving from $\sim0.2$~Hz to $\sim5$~Hz could be related to
the transition radius changing from $\sim 570 R_\mathrm{Sch}$ to $\sim
67 R_\mathrm{Sch}$.  When the disk inner radius gets small enough,
Compton cooling becomes important and the hard Comptonized emission is
suppressed while the black-body emission from the disk moves into the
X-ray band, causing a spectral state transition.  This agrees
remarkably well with what was observed in this outburst.

In order for a two flow model to explain the spectral characteristics
described in \S~\ref{spectev}, it should allow for spectral softening
to occur while the overall source flux is dropping.
For two
independent flows, this type of spectral evolution
occurs because changes in the radial
flow may occur on a short (free-fall) timescale, while the effect
of changes in the the disk accretion rate are delayed due to viscosity
\citep{ct95,shs+01}. 
A model of two independent flows was applied to
GRS~1758$-$258 and 1E~1740.7$-$2942 by \citet{shs+01}, both of which
also showed spectral softening with decreasing flux.  Using a
dependent flow model, \citet{van01} suggests that if the count rate is
dominated by the disk accretion rate and the spectral hardness is
related to the inner disk radius (which is determined by the
time-averaged accretion rate), the same cyclic structure in the HID
will result.

Recent ideas on the role of jets in microquasars suggest a link
between the jet and the corona (see \citealt{fen01} and references
therein) and provide a theoretical basis for the observed association
between the HB and the radio and optical/IR signals
(\S~\ref{sec-multi}).  This association is suggestive of a mechanism
by which jets are created in the LS and quenched in the IS.  The LS is
associated with a continuous radio-emitting outflow and the presence
of the strong LFQPO. At the transition to the IS, discrete radio
ejections seem to be common but the continuous jet ceases and the
LFQPO mechanism is suppressed.  This indicates that both the outflow
and the LFQPO are characteristic of the LS and are related to the
presence of a hot corona or sub-Keplerian flow.

\acknowledgements

We gratefully acknowledge useful discussions with Lev Titarchuk.  
We thank Mark Yashar for providing useful references.
Work at SLAC was supported by department of Energy contract
DE-AC03-76SF00515. Basic research in X-ray Astronomy at the Naval
Research Laboratory is supported by ONR/NRL.  This work was performed
while RMB held a National Research Council Research Associateship
Award at NRL. JDS is grateful to the NASA Applied Information
Technology Research Program for support.  This paper made use of
quick-look results provided by the ASM/RXTE team (see
\url{http://xte.mit.edu}).

\end{document}